# K. Alex Müller: Science between ferroelectricity and superconductivity

R. K. Kremer[1], A. Bussmann-Holder[1], and H. Keller[2]

[1]Max-Planck-Institute for Solid State Research, Heisenbergstr. 1, D-70569 Stuttgart, Germany

[2]Physik-Institut der Universität Zürich, Winterthurerstr. 190, CH-8057 Zürich, Switzerland

K. Alex Müller started his scientific career in 1958 when he was about thirty-one years old. After his wife passed away and being in his nineties, his interest in physics gradually faded. In those years shortly before he passed away, on January 9, 2023, he was no longer interested in superconductivity or ferroelectricity, he had become essentially devoted to philosophy and psychology. It is of note that altogether his research in physics comprised sixty years of activity. Almost in the middle of this period, he was awarded the Nobel Prize in Physics for the discovery of high-temperature superconductivity in ceramic copper oxides in 1987. This discovery made him truly famous. While most researchers are familiar with his Nobel Prize work, many are not aware of the fact that K. Alex Müller was already an acclaimed scientist before that and highly respected in the field of perovskite oxides, ferroelectricity, structural phase transitions, Jahn-Teller physics, electron paramagnetic resonance (EPR) and, especially so in the properties of $SrTiO_3$.

In this contribution, we will review his activities prior to the discovery of the high-temperature superconductivity while providing the evident link to the finding of it. We emphasize specifically that his breakthrough discovery was not accidental. It was predetermined by his deep understanding and wide knowledge of perovskite oxides. In addition, we round out the above overview on K. Alex Müller's pre-Nobel-Prize activities with a scientometric analysis of the publications dealing with the properties of perovskite oxides and especially $SrTiO_3$.

The PhD thesis of Alex was on the EPR investigation on $Fe^{3+}$ in the perovskite $SrTiO_3$ (STO) published in Helvetica Physica Acta in 1958 [1]. The archetype of the perovskite materials is $CaTiO_3$, first described by the German mineralogist Heinrich Rose [2] and named after the Russian mineralogist Count Lev Alekseyevich Perowski, is an oxide that contains metal and oxygen ions in a ratio 2:3. As a mineral ("Tausonite"), STO, the homolog of $CaTiO_3$ replacing Ca by Sr, was discovered only in the early 1980s, whereas the growth of synthetic crystals of STO of good quality to allow the characterization of its physical properties, started already in the early 1950s. That was shortly after the compound had been discovered [3]. In contrast to $CaTiO_3$, STO crystallizes in a cubic crystal structure at room temperature (see Figure 1) and undergoes a structural phase transition to a tetragonal structure at ~105K [4]. In his thesis [1] K. Alex Müller summarizes his microwave resonance experiments on spin S=5/2 ions substituted for Ti in STO. He constructed an X-band superheterodyne EPR spectrometer to measure the resonance of minute traces of the $Fe^{3+}$ ions above and below the structural phase transition in STO. The EPR spectra reflect the fine structure splitting of the 6-fold $S= 5/2$ manifold which change their principal axes when entering into the tetragonal phase.

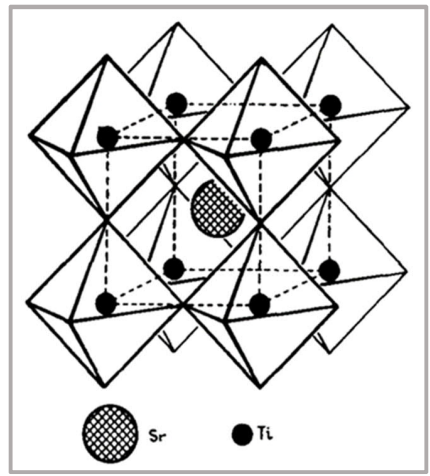

**Figure 1** Crystal structure of STO. Figure taken from the PhD Thesis of K. Alex Müller [1]

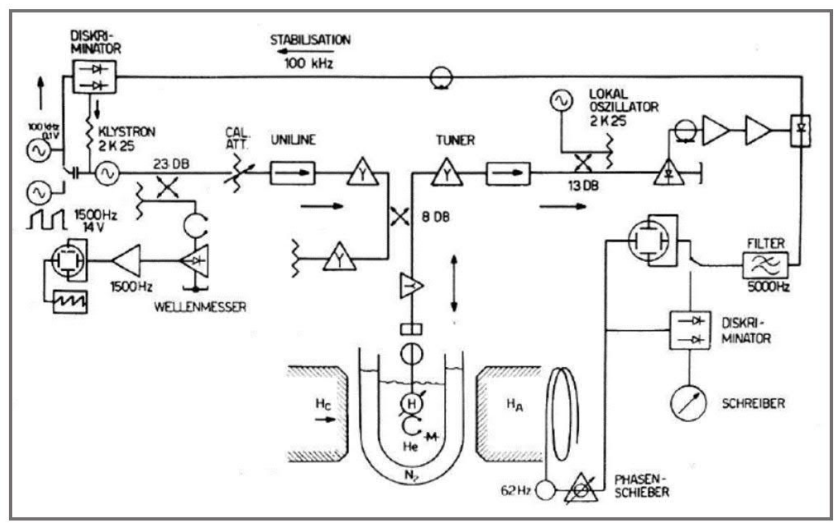

**Figure 2** Scheme of the EPR spectrometer K. Alex Müller constructed in his PhD thesis to measure the resonance of $Fe^{3+}$ in $SrTiO_3$ at room temperature and at liquid nitrogen temperature [1].

At that time, a bit more than a decade after Evgenij Konstantinovič Zavojskij from the Kazan Federal University (Russia) had discovered EPR in concentrated transition metal salts [5], the level of sophistication and perfection of Müller's EPR spectrometer (compare Figures 2, 3) is impressive. Interestingly, the low-temperature dielectric properties of STO which attracted K. Alex Müller's attention about twenty years later, already show up in the concluding remark in his PhD thesis where he wrote: 'Die sehr hohe Temperaturabhängigkeit der DK bei 4,2 °K und die damit verbundene Instabilität der Kavität bei dieser Temperatur verunmöglichte, selbst mit einer Probe von nur 2/10 mm Dicke, eine Messung. Dadurch war eine Bestimmung des Vorzeichens von 3a, durch den Intensitätsunterschied der Linien +5/2→+3/2 und -5/2→ -3/2, und somit auch von D, nicht möglich.'[1]  *[Even with a sample of only 2/10 mm thickness, the very high temperature dependence of the dielectric constant at 4.2K and the associated instability of the cavity at this temperature prevented the determination of the sign of 3a and D from the intensity difference of the resonances lines +5/2→+3/2 und -5/2→ -3/2.]*

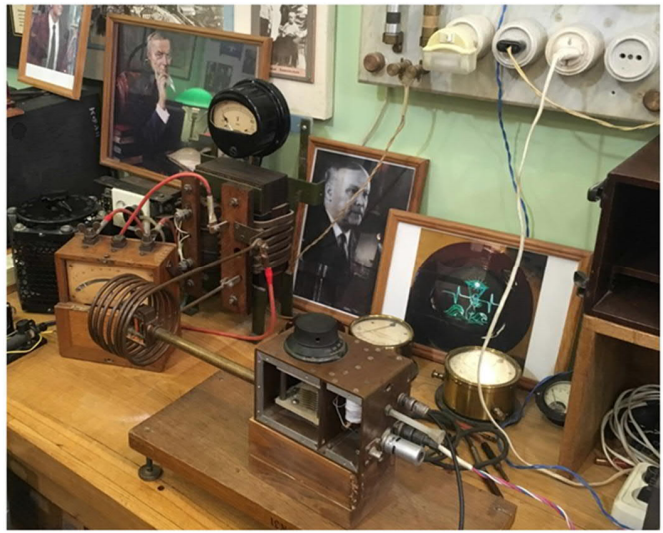

**Figure 3** Restored experimental electron paramagnetic resonance spectrometer of Evgenij Konstantinovič Zavojskij working at a frequency of ~150 MHz displayed at the E. K. Zavojskij Laboratory Museum at the Kazan Federal University. The magnetic field-generating 7-turn coil and the high-frequency generator are seen in the foreground. (private picture taken by R.K.K.).

Shortly after his PhD thesis, K. Alex Müller published a paper on $Mn^{4+}$ impurities in STO in Physical Review Letters [6]. From then on, STO played a central role to Alex and his research paved the way to the nowadays important role which STO has gained in applications as well as in high-end research. His favorite experimental tool was EPR. As compared to long wave-length testing tools EPR is a local probe from which much information on the local environment, crystalline surroundings and structural properties of specific ions can be favorably gained.

The first work on STO was consequently continued by K. Alex Müller. This confirmed for the first time unambiguously the structural cubic to tetragonal phase transition at around 105 K. From thereon, he focused on this zone boundary related instability in oxide perovskites and showed that it is driven by the tetragonal rotation of $TiO_6$ octahedra in STO and the trigonal rotation of $AlO_6$ octahedra in $LaAlO_3$ below their respective phase transition temperatures [7]. The normalized rotation angle φ vary quantitatively in the same way as a function of the reduced temperature in both compounds and φ is the order parameter of such structural phase transitions. However, not only do these two systems exhibit this instability, but it is rather common to most perovskite oxides.

Another scientific highlight of K. Alex Müller and Walter Berlinger [8] is the experimental evaluation of the static critical exponent β of the order parameter φ of the second-order structural phase transition of STO and $LaAlO_3$. Just below the transition temperature $T_S$, the order parameter is described by the power law: $\varphi(t) \sim (1-t)^\beta$ where $t=T/T_S$ and β is the critical exponent. For $t > 0.9$ the order parameter φ(t) is well reproduced by a critical exponent β=0.33(2), indicating that the critical behavior of φ is *universal.* (see Figure 4). However, for $0.7 < t < 0.9$ the order parameter φ(*t*) is consistently described by the mean-field Landau theory with β=0.5. This pioneering work is a true milestone in the field of phase transitions and critical phenomena.

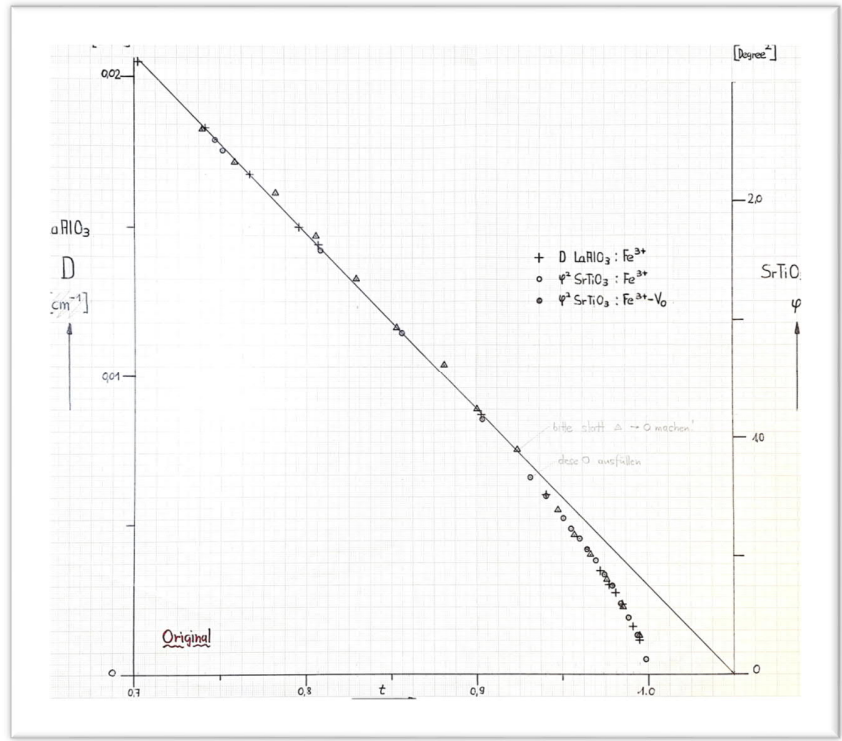

**Figure 4** $\varphi^2(t)$ of STO and crystal field parameter $D$ of LaAlO$_3$ versus reduced temperature $t = T/T_S$ in the range $0.7 < t < 1.0$, showing the changeover from mean-field Landau to critical behavior at $t \simeq 0.9$. This is the original hand drawn figure by K. Alex Müller and Walter Berlinger published in Ref. [8], [private picture taken by H. K.]

Already in 1964, K. Alex Müller published together with Ulrich Höchli his first paper on the Jahn-Teller effect with an investigation on the line broadening of octahedrally coordinated $d^7$ ions with $t^6e$ configuration [9]. The broadening is due to an Orbach relaxation process. For Pt$^{3+}$ in Al$_2$O$_3$ and for Ni$^{3+}$ in SrTiO$_3$ the obtained energy separation $\Delta$ between the lowest two levels was shown to be the Jahn-Teller (JT) splitting of the $^2$E ground state. Relaxation between the two states of a Kramers doublet by the Orbach process takes place when there is an excited energy level located within the phonon spectrum. The relaxation time T$_1$ for this process obeys the relation T$_1$=$A$exp($\Delta$/kT) where $\Delta$ is the separation of the first excited level from the ground doublet and $A$ is the lifetime of the excited state. Since the rather large observed splitting would require three phonon processes to be involved, they suggested the presence of a dynamic Jahn-Teller effect. By analyzing the motion of the JT ion and that of its surrounding ions within the approach of Slonczewski [10], they proposed that an elastically and a centrifugally stabilized mode should occur which lies well above the maximum lattice vibration frequencies.

Besides the zone boundary instability, K. Alex Müller was also interested in the q=0 soft transverse optic phonon modes which are suggestive of a polar instability and a ferroelectric phase transition. The existence of such a mode has been demonstrated in many oxide perovskites and a true transition had been observed in most of them [11]. In addition, the mode softening is indicative of a displacive mechanism in contrast to an order/disorder transition and thereby a classification scheme for these compounds. Since in STO and also other related systems such a mode has been observed, K. Alex Müller together with H. Burkard [12] were curious to find a ferroelectric transition in STO with its very pronounced zone center transverse optic mode softening. A straightforward measurement is the one of the temperature dependence of the dielectric constant ε, which via the Lyddane-Sachs-Teller (LST) relation, is directly linked to the soft mode. Using the LST relation for ferroelectrics the dielectric constant should peak at the transition temperature and follow a Curie-Weiss relation above and below the transition temperature. Such a behavior was anticipated by K. Alex Müller and Burkard when measuring ε($T$) of STO. Instead, they observed a regime at low temperatures where ε($T$) apparently saturated up to the lowest temperatures they could achieve. They figured out that in this low temperature regime quantum fluctuations grow and get larger than the soft mode related displacements to govern completely the dynamics. Accordingly, they named STO a "quantum paraelectric" and defined for the first time this concept.

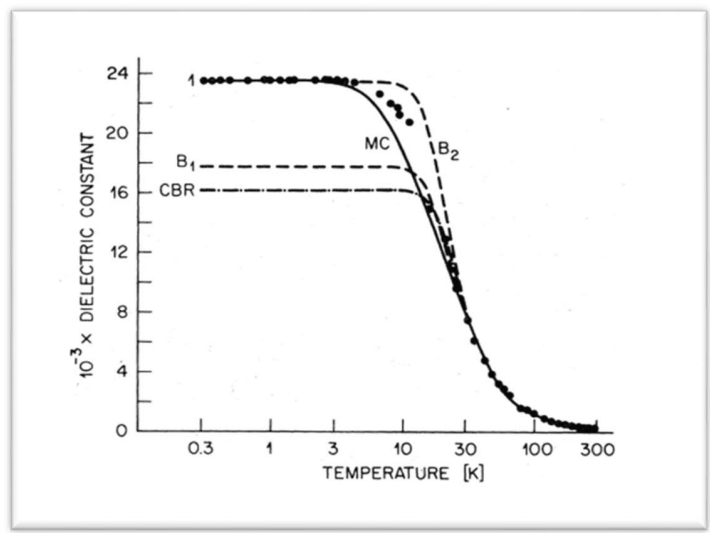

**Figure 5** Dielectric constant of STO in comparison to different theoretical models, for details see [12].

Later, this behavior was observed in many more, almost ferroelectric systems and gained giant attention even nowadays. We emphasize the Müller-Burkard (MB) [12] work especially because, after a decade of moderate interest, it is nowadays extremely often cited with increasing incidence over the years and is currently the highest cited one in the field of quantum paraelectrics (see Figure 6).

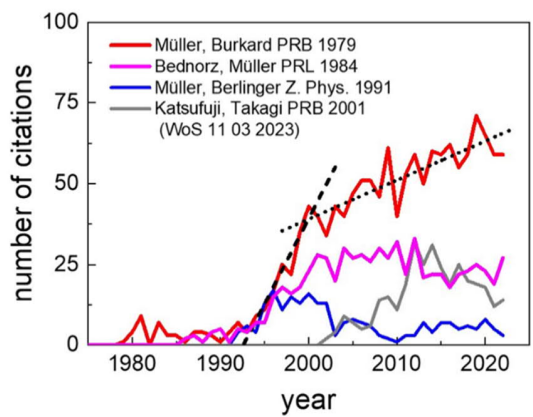

**Figure 6** Number of citations of the four highest cited publications in the research field of quantum paraelectric systems [12, 13, 14, 15]. The dashed straight line corresponds to an increase of 5.3 and the dotted line to a growth of 1.2 citations per year. (Clarivate Web of Science March 11[th], 2023)

In scientometric research, reasons why some publications receive 'delayed recognition' are a matter of broad debate [16]. For example, resistance of the community to new ideas or poor communication skills of the author(s) can be the origin for reduced attention. Causes for a sudden increase in the interest of a publication after a period of delay can be manifold, for example, in the meantime the authors became famous with another scientific achievement as e.g., happened to K. Alex Müller who has been awarded the Nobel prize in 1987. However, significant increase of the citations of the MB paper [12] takes place when interest in high-temperature superconductivity (HTSC) is already on the wane, as is implied by the citation profile of the Bednorz-Müller original HTSC publication [17] of 1986. Figure 7 compares the profiles of the Bednorz-Müller [17] and the Müller-Burkard paper [12], highlighting that the latter gradually began to draw attention only in the middle of the 1990s, about ten years after the discovery of HTSC.

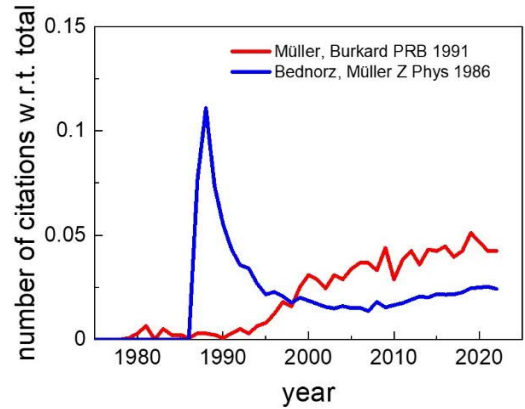

**Figure 7** Number of citations of the Müller Burkard paper [12] in comparison with the original HTSC publication by Bednorz and Müller [17]. The numbers of citations have been normalized to the sum of all citations these papers have accumulated over the years since their appearance. (Clarivate Web of Science March 11$^{th}$, 2023)

What also appears to have supported attention to the MB publication is that the second publication (MBT) [14] on the quantum paraelectric STO appeared in 1991. It is co-authored by Walter Berlinger and Erio Tosatti from Trieste and uses the EPR spectroscopy on $Fe^{3+}$ to investigate the quantum paraelectric state in STO. With respect to the number of citations, this second publication is only surpassed by a report on quantum paraelectricity in $EuTiO_3$ (ETO), a system isotypic to STO, by Katsufuji and Takagi [15] (see Figure 4). However, it is important to note that ETO is not a true quantum paraelectric, since its extrapolated phase transition temperature is around -175K, whereas that of STO is +32K [18]. Nevertheless, $EuTiO_3$ is particularly intriguing because it introduces magnetic degrees of freedom of the half-filled 4f-shell of the $Eu^{2+}$ cations [19].

It is also worthy of mention that a special tribute was devoted to the MB paper [12] by the Editors of Physical Review two years ago [20] as one of the "Physical Review B 50th Anniversary Milestones that have made lasting contributions to condensed matter physics". In addition, the editors had drawn attention to K. Alex Müller's later enormous successes in high-temperature superconductivity with the sentence: *"Müller's deep understanding of the promise of perovskites such as SrTiO₃ led to the discovery with J. G. Bednorz in 1986 of a new class of superconductors, which earned them the Nobel Prize for Physics in the year 1987"*.

While the MB work [12] provided clear evidence that STO would not become ferroelectric at ambient conditions, K. Alex Müller worked together with J. Georg Bednorz on a method on how to achieve polarity in this material [13]. While he had previously concentrated mostly on the role of the transition metal and its chemical bonding, they both investigated the possibility to exchange the A site ion in $ABO_3$. Instead of Sr, they partially substituted the much smaller Ca ion, however only by very small amounts. In the tetragonal phase, the dielectric constant perpendicular to the c axis becomes peaked above $x_c=0.0018$, the quantum mechanical onset of displacive ferroelectricity. The polarization in the *ab* plane can be switched between the two-equivalent *a, b* axes, i.e., the system is an XY, n=2, quantum ferroelectric. Above $x_r=0.016(2)$, $\varepsilon(T)$ peaks round in a special way which is attributed to the onset of a random-field-induced domain state. At x=0.016, the transition becomes diffusive, and the transition temperature range is very broad which originates either from structural disorder or compositional fluctuations. For even larger values of x, the broad peak is independent of the composition and seems to be induced by random electric fields introduced by the $Ca^{2+}$ substitution which has the same charge as $Sr^{2+}$. However, with a much smaller ionic radius than $Sr^{2+}$, it thereby is inducing a random strain, which couples to the polarization. Thus, $Sr_{1-x}Ca_xTiO_3$ mixed-crystals exhibit a transition from XY-type quantum ferroelectricity to a paraelectric parent phase with a transition to a random phase. The related paper has attained considerable attention and is also one of K. Alex Müller's highly cited works (see Table 1). Note, that the work discussed above does not necessarily belong to the list shown in Table 1, but were of importance to K. Alex Müller.

**Table 1** List of the then highest cited publications (number of citations) of K. Alex Müller dealing with the properties of $SrTiO_3$. (Clarivate Web of Science May 1st, 2023)

| Publication | Citations |
|---|---|
| $SrTiO_3$: An intrinsic quantum paraelectric below 4K<br>*K.A. Müller, H. Burkard*<br>Phys. Rev. B **19,** 3593 (1979) | 1400 |
| $Sr_{1-x}Ca_xTiO_3$: An XY quantum ferroelectric with transition to randomness<br>*J.G. Bednorz, K.A. Müller*<br>Phys. Rev. Lett. **52**, 2289 (1984) | 690 |
| Indication for a novel phase in the quantum paraelectric regime of $SrTiO_3$<br>*K.A. Müller, W. Berlinger, E. Tosatti*<br>Z. Phys. B: Cond. Mat. **84**, 277 (1991) | 227 |
| Strong axial electron paramagnetic resonance spectrum of $Fe^{+3}$ in $SrTiO_3$ due to nearest neighbor charge compensation<br>*E.S. Kirkpatrick, R.S. Rubins, K.A. Müller*<br>Phys. Rev. **135**, A86 (1964) | 172 |
| Electron paramagnetic resonance of manganese IV in $SrTiO_3$<br>*K.A. Müller*<br>Phys. Rev. Lett. **2**, 341 (1959) | 153 |
| Electron paramagnetic resonance of three manganese centers in reduced $SrTiO_3$<br>*R.A. Serway, W. Berlinger, K.A. Müller, R.W. Collins*<br>Phys. Rev. B **16**, 4761 (1977) | 109 |
| Observation of two charged states of a nickel-oxygen vacancy pair in $SrTiO_3$ by paramagnetic resonance<br>*K.A. Müller, W. Berlinger*<br>Phys. Rev. **186**, 361 (1969) | 106 |
| Order parameter and phase transitions of stressed $SrTiO_3$<br>*K.A. Müller, W. Berlinger, J.C. Slonczewski*<br>Phys. Rev. Lett. **25**, 734 (1970) | 92 |
| Trigonal to tetragonal transition in stressed $SrTiO_3$: Realization of three-state Potts model<br>*Amnon Aharony, K.A. Müller, W. Berlinger*<br>Phys. Rev. Lett. **38**, 33 (1977) | 91 |
| Photochromic $Fe^{5+}$ ($3d^3$) in $SrTiO_3$ - evidence from paramagnetic resonance<br>*K.A. Müller, Th. von Waldkirch, W. Berlinger, B.W. Faughnan*<br>Solid State Commun. **9**, 1097 (1971) | 75 |

Special attention was devoted by K. Alex Müller to the phase transition sequence observed in $BaTiO_3$ (BTO) where the first transition from cubic to tetragonal at 408K is accompanied by classical soft mode behavior, indicative of a pure displacive transition [21]. At 278 K, BTO becomes orthorhombic, and at 183 K, a transition into the rhombohedral phase takes place. From the EPR measurements performed on $Mn^{4+}$, $Cr^{3+}$, and $Fe^{3+}$ doped BTO, the purely displacive character of these transitions was questioned [22]. Especially in the low-temperature rhombohedral phase, it was observed that $Mn^{4+}$, which substitutes the isovalent $Ti^{4+}$, is displaced off-center by 0.14 Å along the pseudo-cubic (111) directions and is cooperatively coupled to the $Ti^{4+}$ and its dynamics. From the disappearance of the $Mn^{4+}$ EPR spectra in the high-temperature phases, it was concluded that a re-orientational hopping of the $Mn^{4+}$ cations between different (111) off-center directions takes place which signals an order-disorder mechanism [23]. However, from a significant line broadening of the spectra, it became clear that two different time scales were observed which indicate the coexistence of a displacive and an order-disorder component.

From the few examples discussed above, it is evident that long before the HTSC discovery, K. Alex Müller was a worldwide expert in the physics of perovskites, ferro- and antiferroelectricity and structural phase transitions. He was the organizer and co-organizer of the international and European conference on ferroelectricity and is well known in the international community in the field. At the age of fifty-three, he concluded that he could have left his steep scientific research career and continue with an administrative job.

However, on a two-year sabbatical stay at the IBM TJW Research Center in Yorktown Heights (USA), K. Alex Müller turned his interest to superconductivity, a field he had avoided before. He especially observed that the search for superconductivity in oxides was a niche and new discoveries scarce, even though a few oxides were known to exhibit rather high superconducting transition temperatures. By looking closer into these few compounds, he discovered that all of them had a very low density of states at the Fermi energy incompatible with their transition temperatures. This placed them outside of conventional BCS theory and demanded a new and very strong pairing mechanism as, e.g., provided by the Jahn-Teller effect [24]. Only three years later he proved the correctness of this idea together with J. Georg Bednorz.

To finish this account of the story of K. Alex Müller, the final figure (Figure 8) shows the synergy of the above, (see below) highlighting the interrelation between K. Alex Müller's early work and the discovery of high-temperature superconductivity.

1212

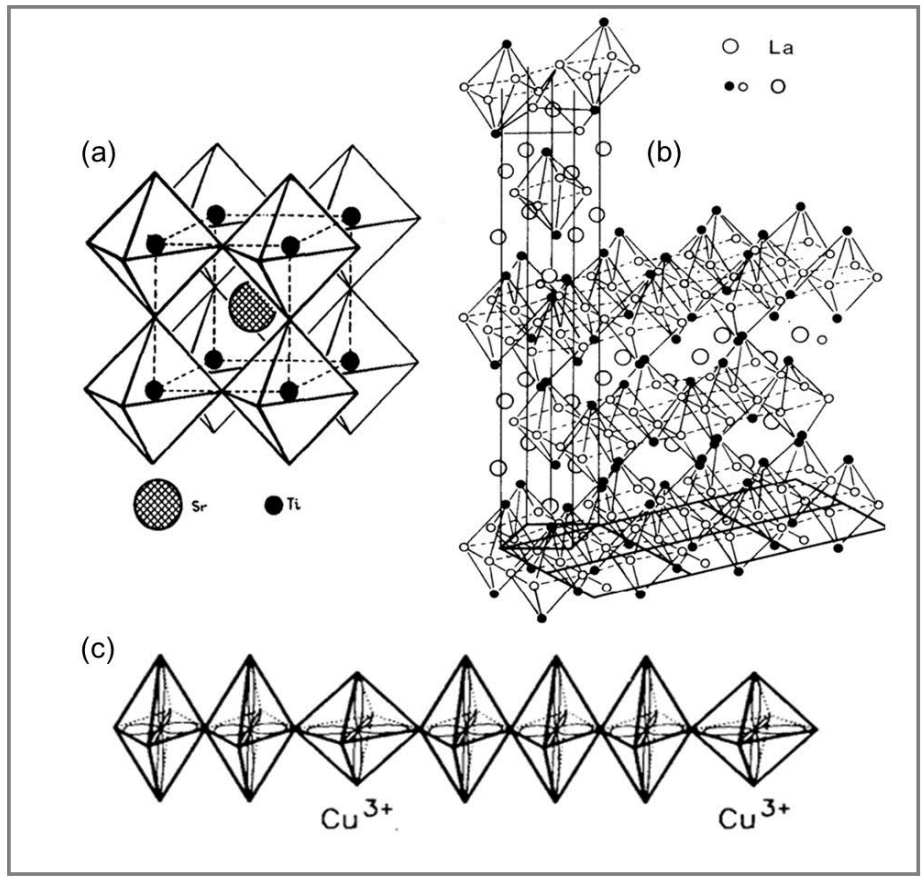

**Figure 8** (a) Crystal structures of SrTiO$_3$ and (b) of La$_{2-x}$Sr$_x$CuO$_4$ and (c) the schematic concept of Jahn-Teller polarons. [25,26].

**Acknowledgement**

We are indebted to É. Reinéry for a careful reading of the manuscript.